New 3D and 2D Octacarbon $C_8$ and isoelectronic $B_4N_4$ having peculiar mechanic and magnetic properties. First-principles identifications.

Samir F Matar*.

Lebanese German University. $C^2M^2S$ (Computer Center for Materials and Molecular Sciences). Sahel-Alma Campus, Jounieh. Lebanon.

*Formerly at University of Bordeaux, ICMCB–CNRS. Pessac. France.

Emails: s.matar@lgu.edu.lb and abouliess@gmail.com

**Abstract**

*Cohesive energies, energy-volume equations of states EOS, electron localization ELF maps, elastic constants, and band structures are reported for original octacarbon C8 three dimensional 3D and two-dimensional 2D chemical systems based on density functional theory calculations. Specifically tetragonal C8 is identified cohesive with hardness close to experimentally identified cubic Ia-3 C8; both exhibiting comparable hardness to diamond. Also isoelectronic and isostructural $B_4N_4$ is calculated with a slightly lower hardness due to the ionocovalent B-N bonding and a bandgap with the same magnitude as diamond. 2D-C8 on the other side is proposed with interpenetrating two carbon hexagonal substructures, identified from energy calculations as stable in a ferromagnetic ground state. Critical pressure for the collapse of magnetization $P_C$=12 GPa let assign a soft ferromagnetic behavior alike Invar alloys. Electronic band structures analyses identify specific bands differentiating magnetic carbon substructure (C1) from nonmagnetic semi-conducting honeycomb-like $C2_6$ layers. These observations let propose spin chemistry perspectives once such multilayered carbon 2D compounds are grown as stand-alone or on selected substrates as thin or thick films.*

Dedication: This paper is dedicated to Professor FR Khalil Chalfoun, Rector of the *La Sagesse* University. Lebanon.

1. Introduction and context

Life and nature are based on carbon as major constituent. But as such, carbon atoms organization into lattices and structures offer a broad range of possibilities regarding dimensionality factor that we focus upon in this work. At zero-dimension 0D, besides fullerene C60, carbon forms into dots entering biosciences applications such as by stabilizing insulin [1]. 1D carbon is mainly illustrated by carbon nanotubes; cf. review by Sinnnott et al. on its syntheses and applications; some of them pertaining to dimensional and chemical compatibility with biomolecules (DNA, proteins, etc.) [2]. Much earlier than technologically elaborated carbon, naturally occurring graphite (2D) and diamond (3D) were known and intensively used. The transformation of graphite into diamond occurs into Earth D'' mantle near volcanoes and tectonic plates, i.e. where the necessary high pressures P >150 kbar and temperatures T ~1200°C are found. However the transformation kinetics is very slow and several millennia are required for obtaining natural diamond. Mimicking Earth P,T conditions, manmade artificial diamond was fabricated in laboratory and General Electric company was the first to produce it in the 1950's [3, 4]. Nevertheless besides its high cost, industrial applications of artificial diamond grafted on tooling machines are hindered by its instability at elevated temperatures generated by the friction involved in the process. Rapidly other substitutes to diamond were sought and boron nitride BN is a manmade binary which is isoelectronic of carbon (2C ≡ B + N) and takes over all its forms, even nannotubes. In the context of present works it is relevant to point out to that BN nanotubes were theoretically devised prior to fabrication by Rubio ate al. back in 1994 [5]. So BN can be synthesized in both 2D *h*-BN (white graphite) and 3D forms: *c*-BN has hardness close to diamond [3] and it usefully substitutes to carbon in tooling machines, due to its better resistance to elevated temperatures with its refractory properties $T_{melt.}$~ 3000°C. Also within the BCN ternary diagram several other artificially devised 2D and 3D new materials were identified by combined experimental and modeling approaches leading to predict new compounds with well targeted properties as extreme hardness (cf. red. [6] for an overview). For instance β-$C_3N_4$ was predicted by Liu and Cohen with a larger hardness than that of diamond [7]. Regarding ternary compounds, $BC_2N$ was predicted in 2001 with larger hardness than c-BN [8] simultaneously with (and independently from) its synthesis [9]. Clearly this research field has benefited (and still is) from theoretical predictions helping subsequent syntheses thanks to accuracy of quantum based calculations within the density functional theory (DFT) [10, 11].

Back to carbon, in 1989 a super dense form, octacarbon C8 [12] was announced as denser than diamond (Fig. 1a). The crystal structure is shown in Fig. 1b and the data in Table 1. The structures feature the characteristic distances which are all close to 1.5 Å. More recently DFT calculations let predict octasilicon $Si_8$ allotrope with potential applications to Li-battery anode materials [13].

The purpose of this work is to presents original structures and results pertaining to electronic structure and energy derived properties for $C_8$ composition and derived isoelectronic $B_4N_4$ structure with original properties resulting from dimensionality, i.e. respectively ultra-hard 3D $C_8$ and $B_4N_4$ and ultra-soft layered 2D C8 ($C1_2C2_6$) formed of interpenetrating two carbon substructures one of them (C1) undergoing magneto-volume effects and long range ferromagnetic order.

## 2. Computational framework

For the sake of fully defining ground structure geometry optimization of the atomic positions and lattice parameters were carried out to minimize the inter-atomic forces close to zero we used the VASP code [14, 15] as based on the projector augmented wave (PAW) method [15, 16] with potentials built within the generalized gradient approximation (GGA) for an account of the effects of exchange and correlation [17]. To relax the atoms in the ground state structure a conjugate-gradient algorithm [18] was used. Structural parameters were considered to be fully relaxed when forces on the atoms were less than 0.02 eV/Å$^3$ and all stress components below 0.003 eV/ Å$^3$. An energy cut-off of 500 eV for the plane wave basis set was used and the tetrahedron method was applied according to Blöchl [19] for both geometry relaxation and total energy calculations. Brillouin-zone (BZ) integrals were approximated using the special k-point sampling of Monkhorst and Pack [20] and successive calculations with increasing precision of BZ integration were carried to reach highest accuracy.

For the relaxed geometry an illustration of the bonding was done through a mapping of electron localization using ELF function [20]. A normalization of the ELF function between 0 and 1 enables analyzing the 2D contour plots following a color code: blue zones for zero localization, red zones for full localization and green zone for ELF= ½, corresponding to a free electron gas. Beside the 2D ELF representation we consider 3D iso-surfaces to further illustrate the offset of ELF between two different chemical species here B-N versus C-C.

## 3. Calculations and results.

### 3.1 Geometry optimization

#### 3.1.1 3D-phases ($C_8$ and $B_4N_4$)

Based on the body centered Si allotrope studied for Li battery electrode needs [13] and considering a primitive cell to allow for selective atom substitutions, $C_8$ was calculated in *P*4/*mmm* N°123 space group with crystal data provided in Table 1a. The C-C distance of 1.51 Å is smaller than in diamond and body centered cubic $C_8$ (cf. Fig. 1). In the simple tetragonal space group the carbon atoms are dispatched into two substructures. In so far that BN is isoelectronic with 2C, we calculated $B_4N_4$ (Fig. 1d) where B and N occupy the two carbon substructures. However this structural hypothesis exhibits presents B-B and N-N unphysical connections besides B-N. A redistribution of B and N within the same plane, i.e. within same 4-fold 4*j* and 4*k* position was operated. The structural setup was then submitted to unconstrained geometry optimization to identify the minimum structure. The calculated data are provided in Table 1b. The resulting structure is shown at Fig. 1c. The larger *a* parameter is due to the larger atomic size of B ($r_B$= 0.82 Å) and the fact that the average ($r_B$ + $r_N$)/2 =0.785 > $r_C$ = 0.770. Effects of change of bonding nature from covalent (only carbon) to iono-covalent -two chemically different species B and N will be illustrated with the ELF plots (Fig. 2).

#### 3.1.2 2D-$C_8$ phase.

Regarding 2D carbon based layer compounds, we recently identified from ab initio the occurrence of magnetic instabilities on nitrogen in carbon rich nitrides as $C_6N$ [21]. The structure belongs to *P*6/*mmm* space group N°191 in a sort of extended $AlB_2$-type hexagonal structure. Also such layered structure is based on experimental evidence of crystal structure of layered carbon based electrodes for Li batteries [22]. The 2D-C8 shown in Fig. 1e consists of two interpenetrating carbon substructures C1 and C2 with 2 C1 at 2.42 Å separation sandwiched between 6 C2 forming a honeycomb network with d(C2–C2) = 1.40 Å. These results arise from calculations done in both non spin polarized (total spins) NSP and spin polarized SP configurations in so far that similar trends towards onset of finite magnetizations were identified in extended carbo-nitride system [21]. The results are shown at Table 1c. While the *a* lattice constant remains almost the same in both NSP and SP configurations leading to similar magnitudes of interatomic distance above, there is a large increase of *c/a* hexagonal ratio from 1.55 to 1.67 pointing to a large increase of volume upon onset of magnetization which amounts to 1.8 $\mu_B$ (i.e. 0.9 $\mu_B$ per C1 atom). The reason for the increase of *c/a* hexagonal parameter and not the *a* parameter, likely arises from the rigidity of the structure involved with the planar carbon honeycomb network in the plane as shown in Fig. 1e in both the magnetic and non structure differentiates from the non magnetic one by the *c/a* ratio increase. The SP total energy being lower than the NSP one by ~0.2 eV (cf. Fig. 4), it can be proposed that the ground state of 2D-C8 is ferromagnetic but close to non-magnetic state due to the low magnitude of energy difference. Lastly an antiferromagnetic order was considered in view of the presence of two magnetic sites with ~1 $\mu_B$ per C1 atom which can align parallel ↑ & ↑ (ferro-) or antiparallel: ↑ & ↓ (antiferro-). Calculations were carried out assigning opposite spins to the two C1's as well as making half the C2 constituents as SPIN UP and the other half SPIN DOWN. The results led to a decrease of magnetic moments down to ~±0.4 and a raise of energy by 0.3 eV, leading confirm the ferromagnetic ground state. However one may expect strong magneto-volume effects which are examined through the establishment of the energy-volume equation of states for the two magnetic configurations in next sections.

### 3.2 Cohesive energies

The difference between total electronic energies and the sum of the constituents' energies allows cohesive energies needed to validate further the geometry optimized structures and establishing trends between related compounds. Table 2 gathers the results for the energies of different atomic constituents and compounds under consideration. The trends can be established upon averaging the cohesive energies as per atomic constituent: $E_{coh.}$/at. Diamond is most cohesive while 2D- $C_8$ is the least cohesive; the other 3D compounds being found in between. Note that the newly proposed tetragonal C8 is more cohesive than body centered cubic C8 (*Ia*-3) and its cohesive energy is closer to that of diamond. $B_4N_4$ shows a cohesive energy close to c-BN ($E_{coh.}$ = -2.52 eV/at. calculated with the same conditions of the title compounds for comparison). This is related with the iono-covalent chemical character arising from the different chemical natures of B (electropositive) and N (electronegative) versus intermediate C, i.e. the difference between C-C and B-N bonding. Such feature is further supported upon visualizing the electron localization in these chemical systems through the Electron Localization Function (ELF).

### 3.3 Electron Localization Function visualization

The ELF 3D isosurface and 2D color maps are shown in Fig. 2 projected over multiples cells for the sake of clear presentation. Relevant to above discussion are the representations of 3D tetragonal $C_8$ and isoelectronic $B_4N_4$ where 3D isosurfaces and 2D slices along vertical diagonal plane are shown to highlight the covalent C-C versus the B-N ionocovalent bonds. Indeed there is an opposite behavior whereby there is a centered torus-like half way between two carbons in $C_8$ signaling a perfectly covalent bond (as in diamond) whereas the grey isosurface has a half-moon shape in $B_4N_4$ pointing to nitrogen signaling an iono-covalent bond polarized away from electropositive boron. Also the tetrahedral coordination can be clearly seen. The opposite natures of the bonds are further illustrated on the right hand side panels with 2D slices showing high localization red areas surrounding the 3D isosurface with clear differentiation between B surrounded with free electron like green areas oppositely to nitrogen surrounded with red strong localization ELF. Turning to $P6/mmm$ hexagonal $C_8$ the 2D and 3D ELF around C1 are different from the honeycomb C2 especially for the half-moon like 3D grey surface on each side signaling non bonded electrons which eventually polarized magnetically; the feature of off-plane electron localization is shown with a slice at z ~0.4 which exhibits red areas signaling strong localization. In spite of the 2.4 Å separation there is significant electron localization of free electron like nature (green ELF ~ ½) showing an non negligible (chemical) interaction between the 2 C1 via σ electrons while the out-of-plane electrons are relevant to π-like; the whole ensemble of valence 2s and 2p C electrons being hybridized in molecular nomenclature as planar $sp^2$. Looking at the slice crossing the C2 substructure, the bonding between adjacent C2 atoms is different with the red localization areas between adjacent C2 atoms. So the two carbon substructures are quite different in electronic and bonding characters.

### 3.4 Energy-volume equations of states and structural stability.

In order to fully establish the ground state (magnetic for 2D-$C_8$) structure, the energy-volume equations of state (EOS) are required. The EOS used here is Birch's [23] expressed up to the 3$^{rd}$ order as:

$$E(V)= E_o(V_o)+[\,9/8]V_oB_o[([(V_o)/V])^{[\,2/3]}-1]^2+[\,9/16]B_o(B^{'}-4)V_o[([(V_o)/V])^{[\,2/3]}-1]^3,$$

where $E_o$, $V_o$, $B_o$ and $B'$ are the equilibrium energy, the volume, the bulk modulus and its pressure derivative. In both cases calculated $B'= 3.6$. The fits are done over (E, V) sets of calculations around minima found from the geometry optimization. Figs. 3 show the E,V curves and fit values in the inserts for the different compounds considered in this work. Focusing on the $B_0$ magnitudes of the 3D compounds, there is a regular decrease from diamond ($B_0$= 420 GPa; 1 GPa = 10 kbar) down to $B_4N_4$ ($B_0$= 350 GPa). $B_4N_4$ is found slightly more compressible than cubic BN with $B_0$ ~396 GPa [24] but it remains a good candidate for ultra-hard material applications, especially with regard to its calculated high cohesive energy. Tetragonal $C_8$ is found highly incompressible with hardness close to that of super dense body centered $C_8$ ($B_0$= 416 GPa). Clearly the argued upon covalent versus iono-covalent characters discussed above and illustrated with the ELF can highlight these results.

Turning to 2D-C8 expressed as $C1_2C2_6$, both NSP and SP EOS were established. The E,V curves are shown in Fig. 4. The SP curve is at slightly lower energy and larger volume due to the development of magnetization, than the NSP curve. It can be observed also that the larger the volume (SP) the lower the bulk module is. The overall magnitude of B0 is very small versus 3D compounds, which is expected in view of the layered nature of 2D-C8 and the large *c/a* ratio. The difference of volumes $\Delta V(SP-NSP) = 8$ Å$^3$ is accompanied by a small energy change $\Delta E(SP-NSP) = -0.15$ eV. This lets suggest that the passage from SP to NSP configuration can be induced by external effects as pressure which can be estimated from the Birch relationship providing a critical pressure $P_C$ [23]:

$$P_C = (B_o/B') [(V_o/V')^{B'}-1]$$

where $B_0$ and $V_0$ correspond to the SP configuration from which one departs to compress the system and $V_1$ the NSP volume. Then one calculates $P_C$ = 12 GPa. This magnitude is much smaller than that required for vanishing magnetization for ferromagnetic oxides as $CrO_2$ ($P_C$ = 150 GPa) as well as intermetallic alloys systems as Fe-Ni ($P_C$ = 55 GPa) [25]. The 2D-$C_8$ can be assigned "soft" magnetic behavior while $CrO_2$ is a "hard" ferromagnet.

3.5 Structural stability.

The elastic properties are determined by performing finite distortions of the lattice and deriving the elastic constants from the strain-stress relationship. In tetragonal and hexagonal symmetries there are six independent elastic stiffness constants $C_{11}$ $C_{12}$ $C_{13}$ $C_{33}$ $C_{44}$ $C_{66}$. The most widely used method of evaluating the elastic stiffness constants is the method of Voigt [26] based on a uniform strain. For the purpose of establishing structural stability of newly proposed 3D and 2D C8 calculations of the respective elastic constants were carried out. The obtained values are given in units of GPa (gigaPascal pressure):

3D-C8 *P4/mmm*:

$C_{11} = C_{22} = 930$; $C_{12} = 170$; $C_{13} = 57$; $C_{33} = 1186$; $C_{44} = 321$; $C_{66} = 446$

3D-$B_4N_4$ distorted from *P4/mmm*:

$C_{11} = C_{22} = 725$; $C_{12} = 167$; $C_{13} = 105$; $C_{33} = 955$; $C_{44} = 241$; $C_{66} = 328$

and 2D-C8 *P6/mmm*:

$C_{11} = C_{22} = 477$; $C_{12} = 117$; $C_{13} = 25$; $C_{33} = 200$; $C_{44} = 180$; $C_{66} = 13$.

As a general trend, all $C_{ij}$ magnitudes are much larger in 3D versus 2D, thus translating at first place the high softness of the latter. Both compounds are structurally stable is so far that the $C_{ij}$ magnitudes are positive while complying with the stability rules pertaining to the mechanical stability:

$C_{11} > C_{12}$; $C_{11}C_{33} > C^2_{13}$; $(C_{11} + C_{12})C_{33} > 2 C^2_{13}$

Quantitatively these $C_{ij}$ magnitudes are rationalized by the calculation of the bulk modulus labeled $B_V$ following Voigt [26]:

$B_V = 1/9 \{2(C_{11} + C_{12}) + 4C_{13} + C_{33}\}$

Numerically, $B_V = 400$ GPa for 3D-$C_8$, $B_V = 340$ GPa for 3D-$B_4N_4$ and $B_V = 159$ GPa for 2D-C8.

These bulk module magnitudes arising from the calculated elastic constants come close to the values obtained from the EOS fit of the E,V curves. This validates the two approaches and confirms the trends announced.

3.6 Electronic band structure calculations.

The electronic band structures of the different compounds were calculated in the respective Brillouin Zones BZ at high *k-* mesh integration. Fig. 3 shows in 4 panels the band structures of the 3D compounds. The zero energy along *y*-axis is with respect to the top of the (occupied) valence band VB ($E_V$) separated from the (empty) conduction band CB by an energy gap, leading to the labeling (E-$E_V$) eV. The *x*-axis spans the main directions of the irreducible respective BZ's.

Starting with diamond the band gap is ~5 eV as commonly found in the literature (cf. [24] and therein cited works). With body centered cubic $C_8$ the band gap decreases to half its magnitude while in tetragonal $C_8$ the gap magnitude increases again up to ~4 eV. In $B_4N_4$ and the separation from CB is very large with the band gap ~5 eV almost the same as in diamond. This is an interesting result, knowing the large chemical differences between C and BN. The bands have the same shape as tetragonal $C_8$ in spite of the small distortion of the lattice due to the atomic rearrangement and the symmetry decrease due to the loss of the 4-fold (*$C_4$*) vertical symmetry axis and its replacement by a 2-fold (*$C_2$*) symmetry axis.

Regarding 2D-C8, Fig. 4 shows the band structures differentiated for the non-magnetic substructure C2 (top) and magnetic C1 (bottom) for the sake of clarity. For the C2 substructure comprising six atoms forming a honeycomb network (Figs. 1 and 2), there is a small band gap as in graphite and the energy reference is with respect to $E_V$ top of the VB. This is opposed to C1 substructure which shows two sets of bands separated by a small energy due to the magnetic polarization whereas C1 bands show splitting into majority spin bands (↑) at lower energy than higher energy bands corresponding to minority spins (↓). The energy split between the two spin populations involves all band even the low energy s bands. This can be assessed within a molecular like scheme whereby there is hybrization of s and p valence state of carbon provide *$sp^2$*-like planar hybridization with one σ electrons connecting the two C1 carbon atoms and (cf. Fig. 2c) and off-plane 2π like. Oppositely to C2 substructure, there is no gap between VB and CB and C1 bands cross the top of VB at $E_F$ (Fermi energy) because of the metallic character characterizing C1 substructure. The corresponding population difference (↑ - ↓) provides the magnetization which amount to 0.9 $\mu_B$/C1. Clearly we are presented with two interpenetrating carbon substructures behaving differently while being in the same 2D structure C8.

### 4. Conclusions.

The purpose of this work is to propose carbon new forms with original properties in 3D and 2D networks identified and characterized with help of quantum mechanical tools. Such protocol has been shown to be pertinent in proposing new light elements (B, C, N) based hard materials for many decades now. Clearly this research field has long benefited from theoretical predictions helping subsequent syntheses thanks to accuracy of quantum based calculations within the density functional theory (DFT). Here ultra-hard tetragonal C8 and isoelectronic $B_4N_4$ have been devised and characterized with energy-volume EOS and electronic band structures as well as electron localization visuals together with ultra-soft 2D-C8 with peculiar magnetic properties found to be volume dependent, thus highlighting magneto-volume effects characterized by a critical pressure for magnetization vanishing. It is hoped that subsequent synthesis efforts as previously done for theoretically devised $C_3N_4$, $BC_2N$ etc. can be paid to prepare such original compounds.

TABLES

Table 1: C$_8$ and B$_4$N$_4$ new forms in 3D and 2D dimensional structures. Experimental and (calculated) data for C$_8$ *Ia*-3 and calculated parameters for other original compounds. The *a* lattice parameter and the interatomic distances in Å.

a) 3D C$_8$ phases

| Space group | C$_8$ *Ia*-3 [4] | C$_8$ *P4/mmm* (*) |
|---|---|---|
| a | 4.29 (4.47) | 4.38 |
| c/a |  | 0.57 |
| Atomic positions |  |  |
| C1 | (16c) 0.104 (0,094), *x, x* | (4j) 0.180,0.180,0 |
| C2 |  | (4k) 0.321,0.321, ½ |
| d(C-C) | 1.54 (1.52) | 1.51 |

b) 3D B$_4$N$_4$ phase distorted-tetragonal from *P4/mmm* with B and N mix- positions of (4j) and (4k).

| *Dist. Tetrag.* | B$_4$N$_4$ |
|---|---|
| *a* | 4.40 |
| *c/a* | 0.58 |
| Atomic positions |  |
| B | 0.174, *x*, 0; 0.826, *x*, 0; 0.675, 1- *x*, ½; 0.325, 1- *x*, ½ |
| N | 0.813, 1- *x*, 0; 0.187, 1- *x*, 0; 0.325, *x*, ½; 0.675, *x*, ½ |
| d(B-N) | 1.53 |

c) 2D C8

| *P6/mmm* | NSP | SP |
|---|---|---|
| a | 4.18 | 4.19 |
| c/a | 1.55 | 1.67 |
| *Total energy (eV)* | -61.37 | -61.48 |
| *Volume* | 97.83 | 111.57 |
|  |  |  |
| C1 | (2d) 1/3, 2/3, ½ | (2d) 1/3, 2/3, ½ |
| C2 | (6j) 0,0.333,0 | (6j) 0,0.333,0 |
| d(C1-C1) | 2.42 | 2.42 |
| d(C2-C2 | 1.39 | 1.40 |
| M ($\mu_B$) | - | 1.80 |

Table 2: Diamond, octacarbon and $B_4N_4$ forms. Trends of cohesive energies ($E_V$). Energy per constituent atom: E(C) = –7.11 eV; E(B)= –5.55 eV; E(N)= –6.82 eV.

| Compound | Space group | $E_{Tot.}$/8FU | $E_{coh.}$/at. |
|---|---|---|---|
| Diamond | *F-43m* | –72.76 | –1.985 |
| 3D-$C_8$ | *Ia-3* | –67.14 | –1.283 |
| 3D-$C_8$ | *P4/mmm* | –71.18 | –1.786 |
| 3D-$B_4N_4$ | distort. tet. | –68.87 | –2.423 |
| 2D-$C_8$ | *P6/mmm* | –61.47 | –0.574 |

# FIGURES

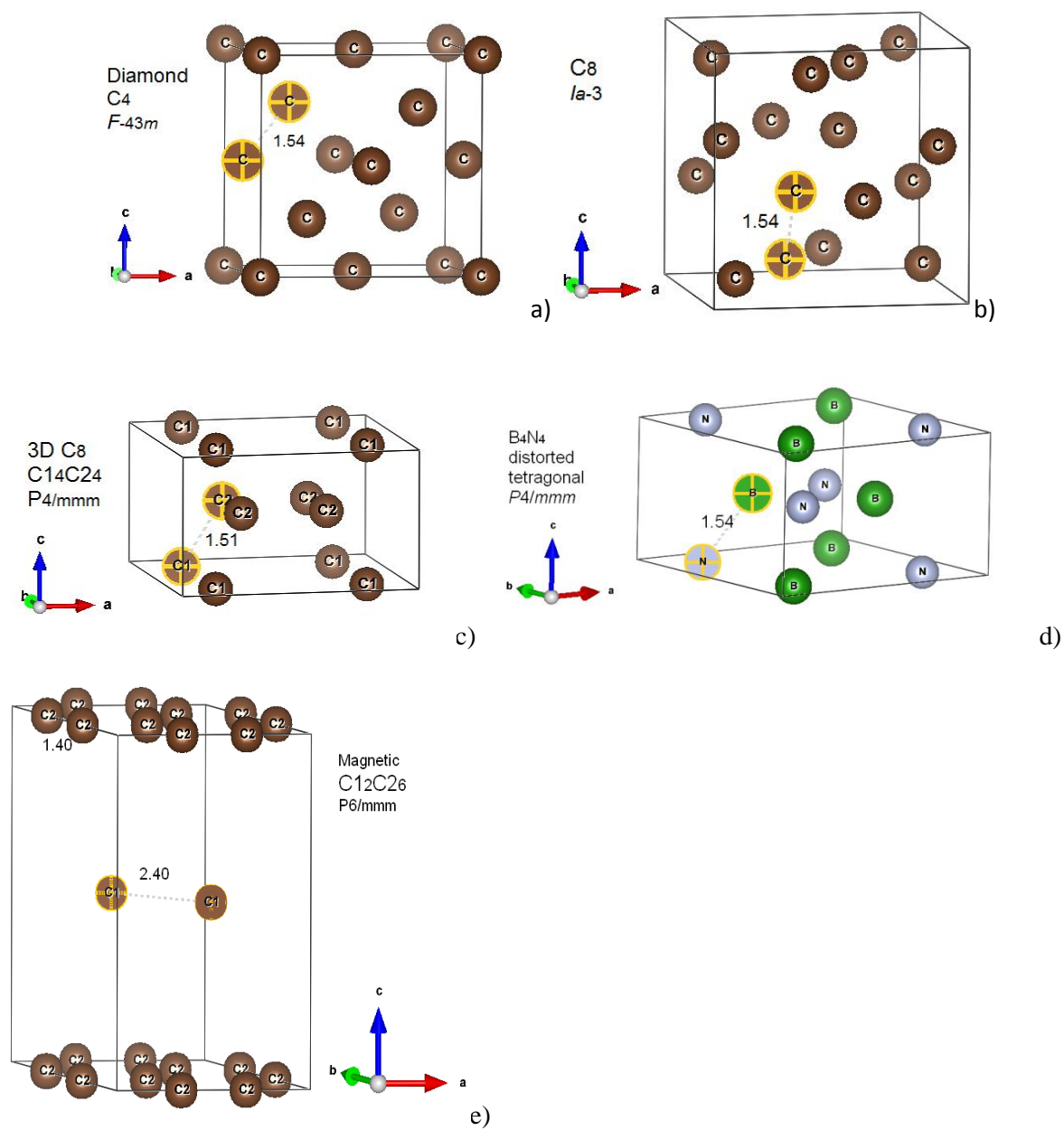

Fig. 1. Crystal structures of the carbon systems (and BN) under consideration: a) Diamond; b) C8 *I-a*3 ; c) $C_8$ *P*4/*mmm* ; d) $B_4N_4$ distorted *P*4/*mmm*; e) Magnetic 2D-$C_8$ *P*6/*mmm*. Remarkable distances in Å are shown.

N.B. Drawings were produced by VESTA ref. K. Momma and F. Izumi, "VESTA 3 for three-dimensional visualization of crystal, volumetric and morphology data," J. Appl. Crystallogr., 44, 1272-1276 (2011).

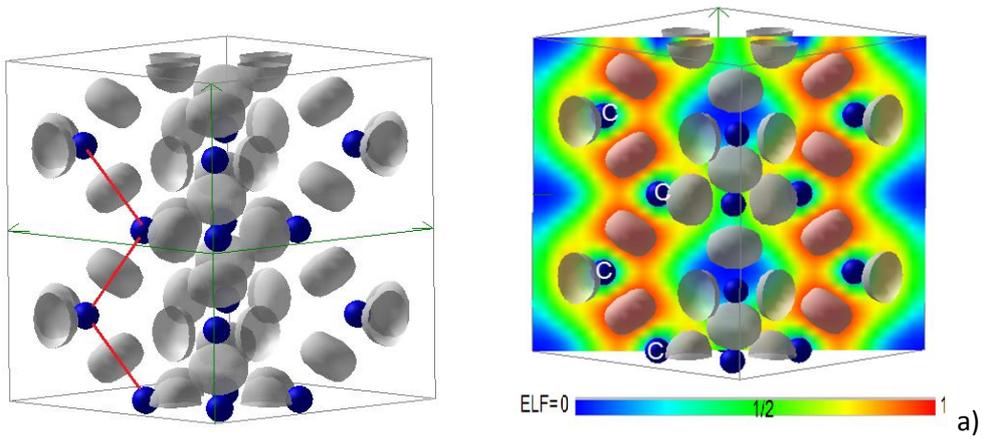

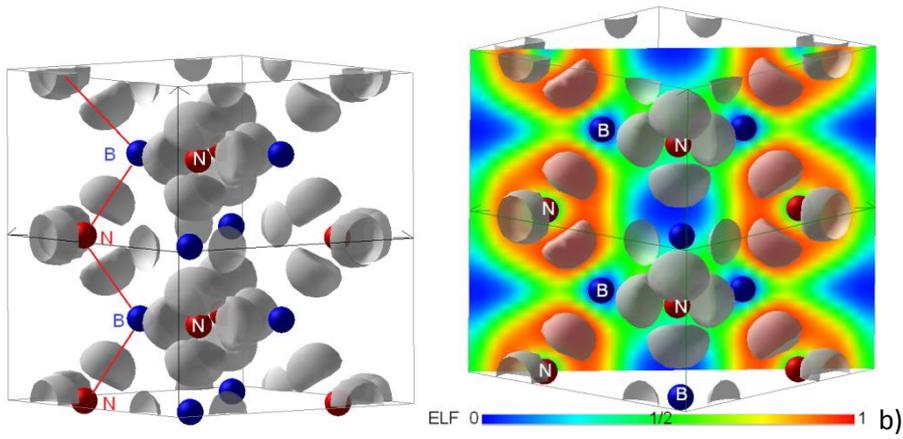

**C1₂C2₆**

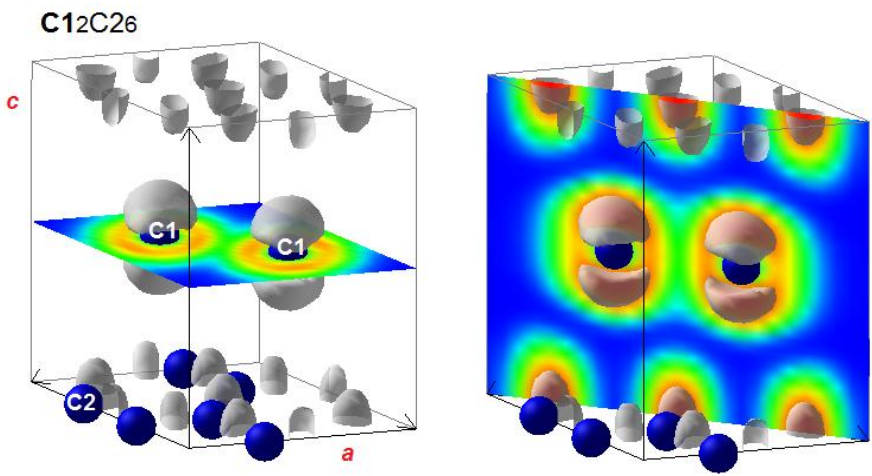

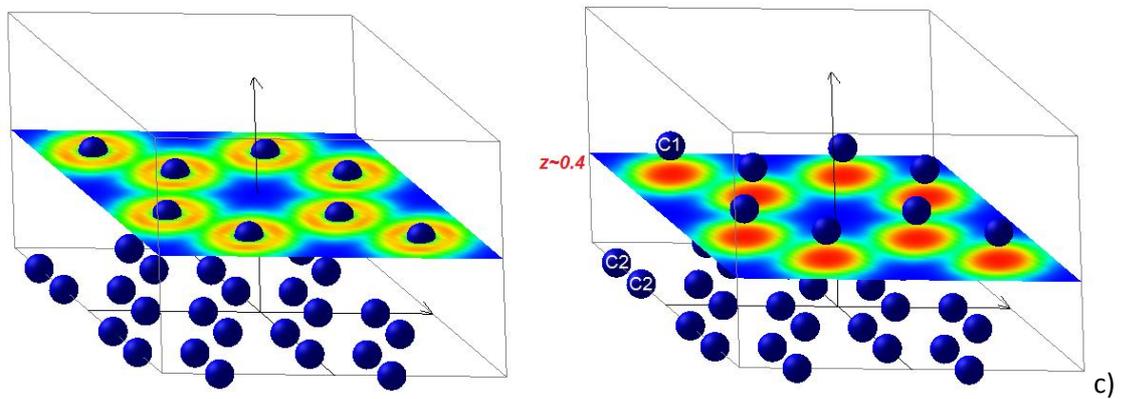

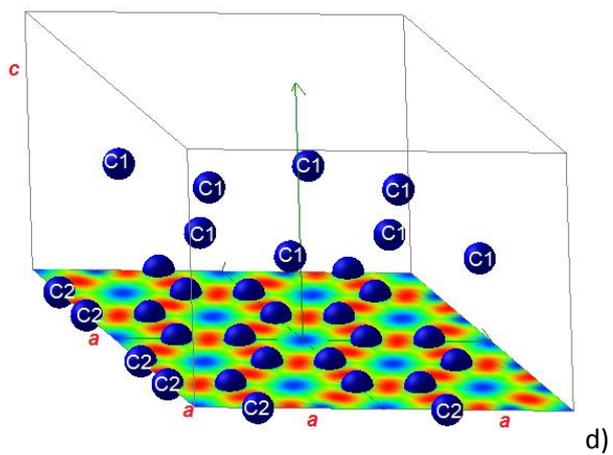

d)

Fig. 2 Electron localization 3D and 2D mapping over multiple adjacent cells in 3D tetragonal C8 (a), in 3D $B_4N_4$ (b) and 2D $C_8$ ($C1_2C2_6$). The ruler is meant to show color code versus ELF magnitude from 0 to 1.

.

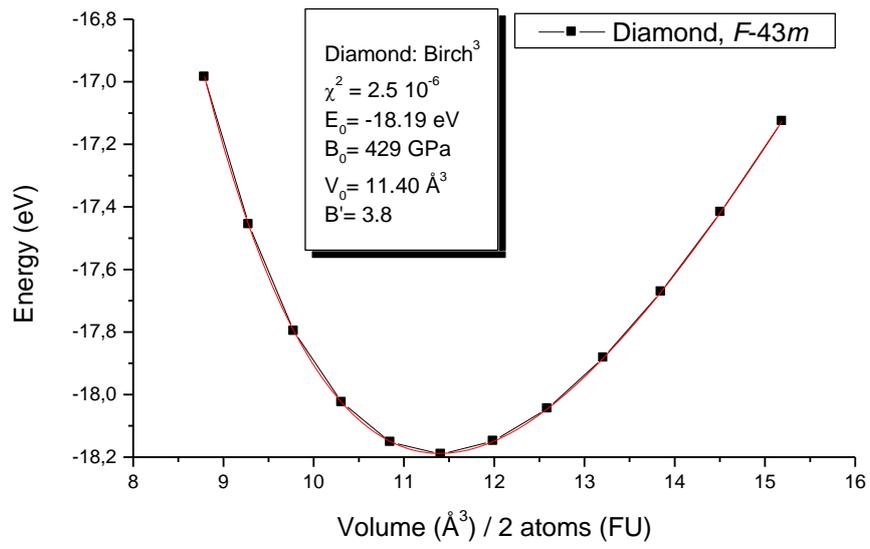

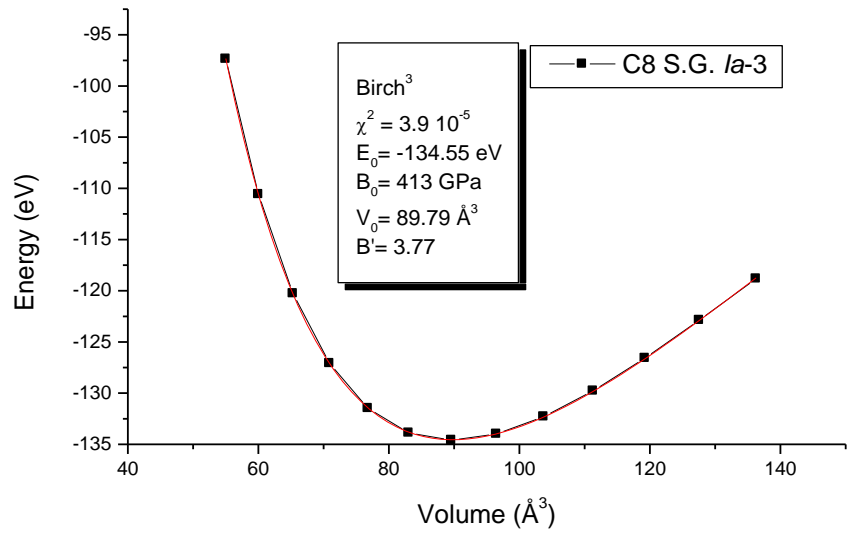

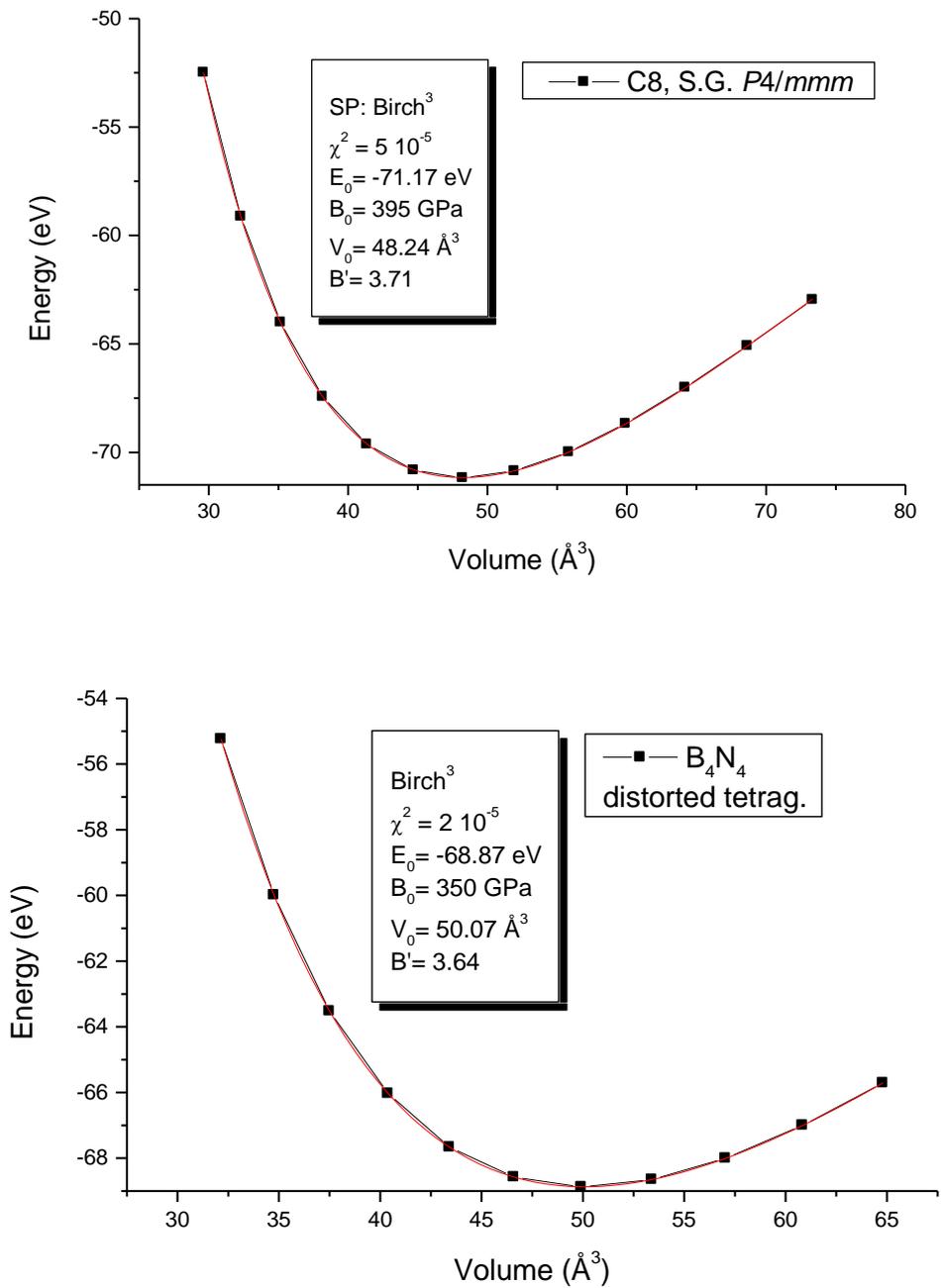

Fig. 3. 3D carbon derived compounds: cubic diamond, body centered C8 and new 3D carbon $C_8$ and tetragonal boron nitride $B_4N_4$ phases: Energy - volume quadratic curves and fit parameters (inserts) of 3$^{rd}$ order equation of states (EOS) (see text)..

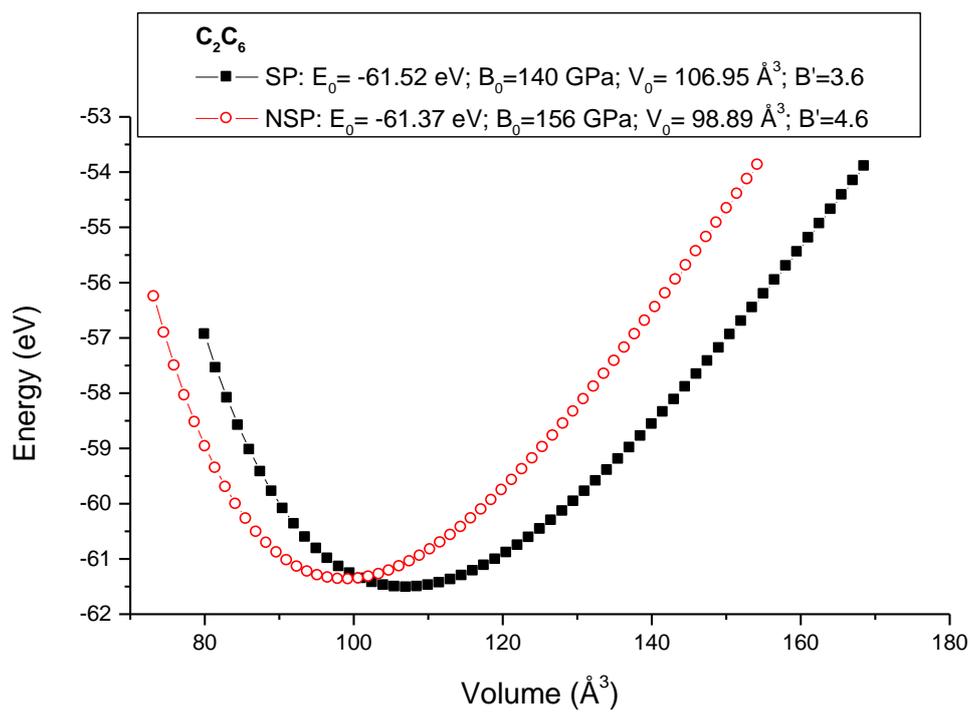

Fig. 4. 2D octacarbon phase: Energy-volume quadratic curves and fit parameters (inserts) of 3$^{rd}$ order EOS.

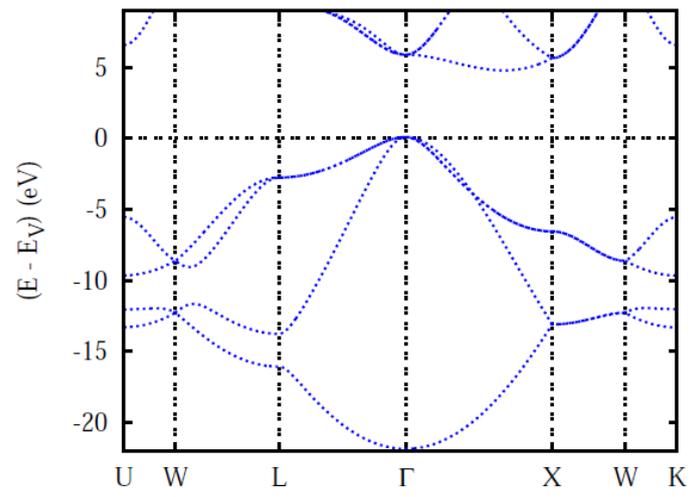

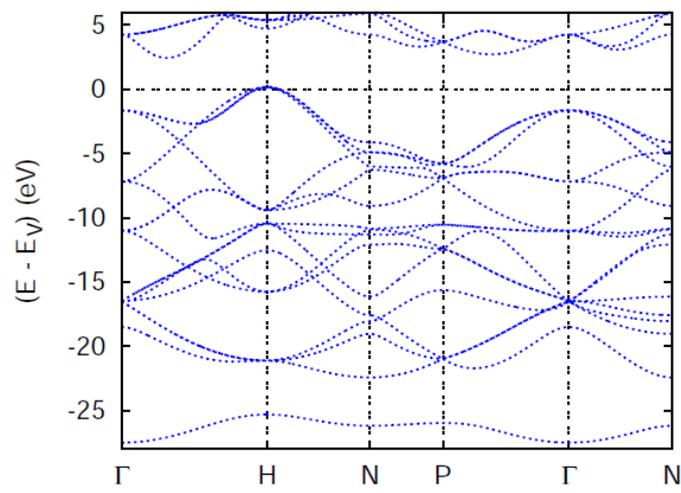

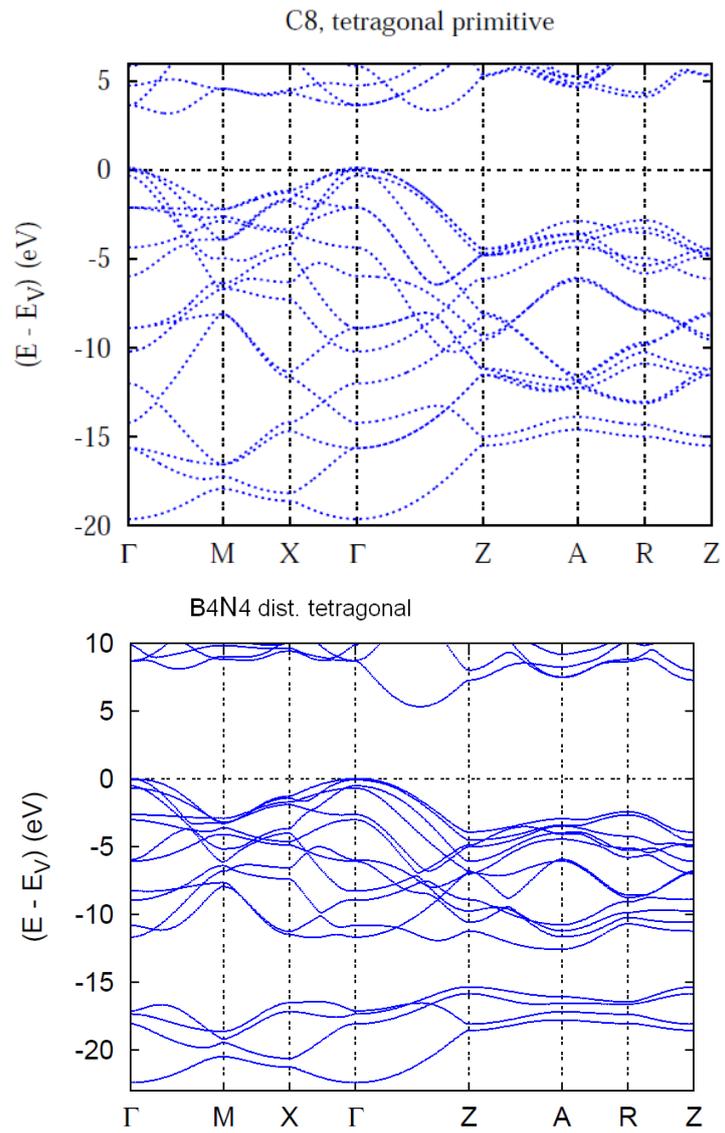

Fig. 3. Electronic band structures of 3D title compounds.

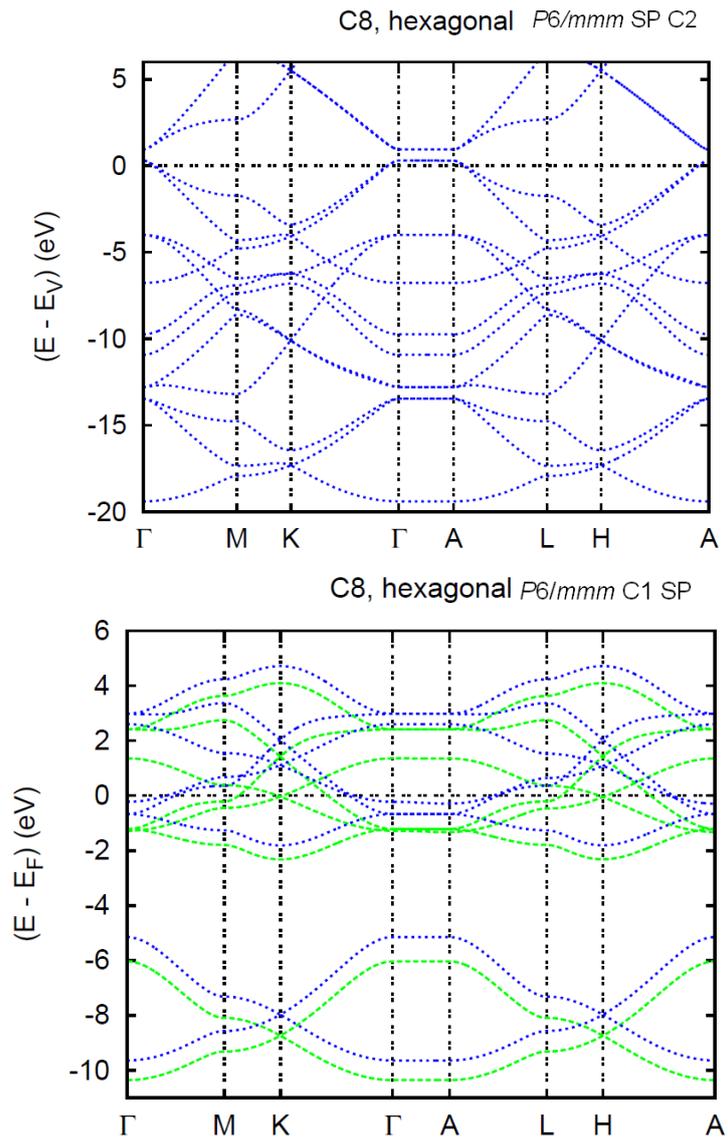

Fig. 4. Electronic band structures of 2D C$_8$ (C1$_2$C2$_6$) in magnetic configuration, highlighting semi-conducting behavior within *C2*$_6$ substructure and spin polarization within **C1**$_2$ bands with a magnetization of 1.8states diamond and the 3D octacarbon phases.